# Edit Transactions

## Dynamically Scoped Change Sets for Controlled Updates in Live Programming

Toni Mattis[a], Patrick Rein[a], and Robert Hirschfeld[a]

a   Hasso Plattner Institute, University of Potsdam, Germany

**Abstract**   Live programming environments enable programmers to edit a running program and obtain immediate feedback on each individual change. The liveness quality is valued by programmers to help work in small steps and continuously add or correct small functionality while maintaining the impression of a direct connection between each edit operation and its manifestation at run-time. Such immediacy may conflict with the desire to perform a combined set of intermediate steps, like a refactoring, without instantly taking effect after each individual edit operation. This becomes important when an incomplete sequence of small-scale changes can easily break the running program.

State-of-the-art solutions focus on retroactive recovery mechanisms, such as debugging or version control. In contrast, we propose a proactive approach: Multiple individual changes to the program are collected in an *Edit Transaction*, which can be made effective if deemed complete. Upon activation, the combined steps become visible together.

*Edit Transactions* are capable of dynamic scoping, allowing a set of changes to be tested in isolation before being extended to the running application. This enables a live programming workflow with full control over change granularity, immediate feedback on tests, delayed effect on the running application, and undos at the right level of granularity.

We present an implementation of *Edit Transactions* along with *Edit-Transaction*-aware tools in Squeak/Smalltalk. We asses this implementation by conducting a case study with and without the new tool support, comparing programming activities, errors, and detours for implementing new functionality in a running simulation. We conclude that workflows using *Edit Transactions* have the potential to increase confidence in a change, reduce potential for run-time errors, and eventually make live programming more predictable and engaging.

**ACM CCS 2012**
- **Software and its engineering → Development frameworks and environments; Object oriented development;**

## The Art, Science, and Engineering of Programming

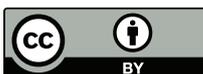

## 1 Introduction

Programming environments can be classified according to the timeliness of run-time feedback. Environments that provide *immediate* feedback for any change at run-time are called *live programming environments* [26]. Examples include *Lively* [13], *Self* [28], some *LISP* environments, *Squeak/Smalltalk* [9, 12], and educational environments like *Scratch* [20], and *Etoys* [8].

Immediate feedback sustains the impression of causality between change of source code and change in run-time behavior, e.g. a previously failed unit test succeeding right after the programmer saves that change, or a button having an effect when clicked after updating its event handler. This immediacy often makes it difficult to work on a more complex change, involving more than a single edit step, such as a refactoring or implementation of a cross-cutting concern, without facing the imminent danger of breaking the running program.

For example, if programmers decide to insert a call to a method that has not been implemented yet, they could do so in non-live environments and implement the missing method later before they run the program. In Smalltalk, for example, the code referring to a planned, but not yet implemented method, may fail at any moment if it is part of the control flow of a running program. This forces programmers to pause their work on the call site, implement the missing method first, and then return to the original method to save the code.

From the perspective of the programming environment, the feedback cycle can be modeled as a two-stage process [19]:

1. *Adaptation*, the phase where a changed translation unit, e.g. a method in Smalltalk, is translated to its executable representation. The feedback programmers obtain in this phase is restricted to compile-time errors.

2. *Emergence*, the phase where the control flow reaches the newly translated program parts and executes them, possibly causing a change in behavior that becomes visible.

In this model of immediate feedback, the problem manifests itself as the fact that the frequency of emergence is directly linked to the granularity of adaptation, i.e. after each method change in Smalltalk, the newly compiled method becomes reachable by the control flow. The following proposal focuses on separating emergence from adaptation to allow programmers to change a desired number of translation units, while giving full control of the time and scope of emergence to programmers.

Most contemporary programming environments are equipped with advanced debugging facilities that enable to fix errors as they occur or revert the program to a previous version that worked, but all of these facilities can only react to a failure already emerging.

### 1.1 Changes as *Edit Transactions*

Rather than retroactively repairing a program after it failed, we propose to proactively prevent changes from becoming automatically and globally effective. We aim at situations where it is clear from the start that multiple locations across the program need to be modified.

Our approach considers program execution and modification as concurrent activities on the program's representation, i.e classes and methods in class-based object-oriented programming, and models them as *Edit Transactions* with atomicity properties, isolation, and scope.

We give programmers explicit control over the scope where *Edit Transactions* are active. By allowing a different composition of active *Edit Transactions* in different control flows, programmers can explore new behavior resulting from recent changes, such as running unit tests, without affecting an already running program.

We present an implementation in Squeak/Smalltalk based on dynamically scoped views on state, methods, and classes, such that running programs are presented a view into the system containing the old representation, but development tools and test runners may already use and give feedback on a future version. Once the *Edit Transaction* is activated globally, every control flow will atomically consider the new program representation.

## 1.2 Case Study

We demonstrate *Edit Transactions* through a case study on a graphical actor-based disease spreading simulation running in Squeak/Smalltalk. We implement the concept of recovering from infections using an *Edit Transaction* based workflow and compare it to a workflow without *Edit Transactions*. We show which types of errors occurred in the two workflows and describe notable differences in the programming practice, such as the order in which methods were changed. We also discuss the impact of *Edit Transactions* on the responsiveness of the programming system.

## 1.3 Structure of this Work

In the following, we first introduce the concept of *Edit Transactions* in section 2. Based on this conceptual description, section 3 demonstrate how *Edit Transactions* can be integrated into the programming workflow through conceptual tools. In section 4 we describe our implementation of *Edit Transactions* for the Squeak/Smalltalk environment which we use to illustrate the concept in the case study documented in section 5. Section 6 points out limitations of existing meta programming facilities we encountered in designing and implementing *Edit Transactions* and outlines future research directions. In section 7 we compare *Edit Transactions* to related approaches and section 8 concludes the paper.

```
1   Simulation >> mainloop
2      [running] whileTrue: [
3         self ball step.
4         self wait: 10]
5
6   Ball >> step
7      self bounce; move; gravitate          ⟸(1)
8
9   Ball >> move
10     self position: (self position + self speed)
11
12  Ball >> gravitate                          ⟸(2)
13     self speed y: (self speed y - Simulation gravity)
14
15  Ball >> bounce [...] Check edges and reflect
```

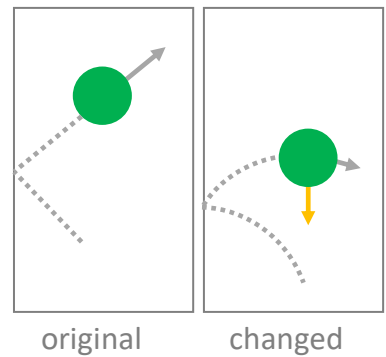

original          changed

■ **Figure 1** Smalltalk code running a bouncing ball simulation. A composite change marked by ⟸ should add gravity. Both changes should become effective at same time, but in Smalltalk, the first change would already take effect after **step** is updated and crash the simulation.

## 2 *Edit Transactions*

### 2.1 Introductory Example

As an introductory example, consider a game that involves a ball moving over the screen and bouncing off the edges of its container. Smalltalk code for this scenario is given in figure 1. As is customary in live programming, the game is constantly running. The next step is to implement gravity via a separate method gravitate that adds acceleration (see figure 1).

In Smalltalk, the granularity of change is a method, which means that updating the step method to call gravitate would take effect immediately and trigger an exception complaining that the message is not understood. Even worse, if there were multiple stepping objects, each object would fail individually and raise an exception. The workarounds are either implementing the method first and then call it, which requires foresight, or halting the simulation, which sacrifices liveness.

We propose a different workflow, in which programmers can tell the editor that a composite change is going to happen, similar to starting a transaction in a database. We call this part *staging*. Then, programmers save the modified step method, which already gives feedback on the syntactical correctness. If auto-testing is active, unit tests would already run and signal that step raises an exception. After implementing gravitate, programmers can *activate* both changes atomically. If the outcome is unexpected, e.g. gravity has the wrong direction, both changes can be *deactivated* immediately. If everything is fine, both changes can be committed to the code base, which we call *merge*.

This concept of combining fine-grained adaptation that is partially *live* with a controlled granularity and scope of emergence will be called *Edit Transaction* and is defined below.

### 2.2 The Concept of *Edit Transactions*

An *Edit Transaction* is defined as a set of changes to a program's *meta-objects* (defined below) with a guarantee that all changes come into effect simultaneously (*atomicity*) and can be undone immediately. The life cycle of a *Edit Transaction* is characterized by the following concepts which can be regarded as operations applicable to the set of changes it represents:

**Staging** While an *Edit Transaction* is *staged*, it collects changes to the program and prevents them from modifying the underlying base system.

**Activation** In a defined *scope*, an arbitrary set of *Edit Transactions* can be *activated*, which emulates the corresponding sequence of composite changes being applied atomically. The scope can be extended over time to other threads and control flows to obtain feedback. Deactivation undoes the changes atomically.

**Merging** If the flexibility of an *Edit Transaction* is not needed any more, it can be *merged* into the base system and become permanent part of the program.

**Aborting** If the captured changes are not needed anymore the *Edit Transaction* can also be *aborted*. This will, if applicable, un-stage, deactivate, and delete the *Edit Transaction* from the system.

Activation, deactivation, and merging are subject to concurrency control to guarantee certain isolation and consistency properties for currently executing programs. Further, while an *Edit Transaction* is staged, the following two operations are possible:

**Add a new version of a meta-object** The change to the meta-object becomes part of the *Edit Transaction*. The new version can also represent a meta-object which does not exist in the unmodified system. There is always only one version per meta-object in a transaction.

**Removing a meta-object** If the transaction includes a version for the meta-object, this version is removed. Otherwise, a removal of a meta-object is recorded in the transaction.

### 2.3 Meta-object Changes

In the context of this work, *meta-objects* are the objects a program consists of, i.e. classes and methods in class-based object-oriented environments. This definition is independent from the meta-objects' representation, e.g. whether they exist as source code files or as first-class object within the running execution environment, because in a live programming environment both should be conceptually the same and properly synchronized to reflect code changes at run-time.

*Edit Transactions* are concerned with capturing changes to meta-objects. In a class-based environment, we consider the following change operations that can be made by editing tools:

- Creation, deletion, renaming of a class
- Changes to the class hierarchy
- Creation, deletion, renaming of a field in a class

- Creation, deletion, renaming of a method

In case the environment makes use of other programming models and modularity concepts, such as packages, actors, layers, roles, etc., those need to be considered as meta-objects as well. This work will focus on class and method modifications.

**Invariants**    When recording changes to an *Edit Transaction*, we need to make sure some invariants are met before an *Edit Transaction* is *valid* and can be activated during program execution:

**Hierarchy Invariant**  Each class has exactly one inheritance path to the top class. A multiple inheritance approach would involve a more sophisticated method dispatch which is beyond the scope of this work.

**Field Invariant**  A field name may appear at most once in the inheritance path from any class to the top class.

While these invariants are concerned with the static program structure and inherent to most object-oriented environments, we impose additional invariants on an *Edit Transaction* during run-time, which are more specific to live programming environments:

**Class Preservation Invariant**  Each object maintains its class over the whole lifetime. That means, a class will remain in the system (possibly anonymous) as long as it has instances after deletion.

**Information Preservation Invariant**  Each object preserves the most recent value assigned to each of its fields independently of activation or deactivation of an *Edit Transaction* removing or adding this particular field. This ensures that an object's field is still readable even if an *Edit Transaction* deleting this field has been active in a different control flow; and guarantees that an assigned value is not lost because the *Edit Transaction* adding this field has temporarily been deactivated.

**Changing methods**    We assume that the smallest unit of change regarding behavior is a *method*, i.e. after updating and saving any code within a method, the executable representation of the whole method is updated. For each modified method, the topmost staged *Edit Transaction* stores both the *executable* representation of the new method and the source code it was compiled from.

Effectively, *Edit Transactions* add another *dimension of dispatch*. When an object receives a message, the default behavior is checking whether the class itself responds to this message, i.e. has a method for it. If not, the lookup recurses into the superclass, eventually finding a method to invoke or raising an appropriate exception after reaching the top of the class hierarchy. With a set of active *Edit Transactions*, the dispatch at each class needs to first look if any of the *Edit Transactions* has a modified method for this class before proceeding to the unmodified class and subsequently traverse the *Edit Transaction* stack again at the superclass.

Since an *Edit Transaction* can contain a method previously not understood by the class, we can trivially add a method by storing it to the *Edit Transaction* under the new message selector. Method removal means to install an indicator in an *Edit Transaction* that the given message needs to be dispatched to the superclass immediately and not to the current class.

If the same method is being updated multiple times within a single *Edit Transaction*, only the most recently compiled version is retained.

**Changing fields**   For each modified class, the staged *Edit Transaction* needs to store which fields have been added and removed. Effectively, activating an *Edit Transaction* overlays all instances of the changed class with a new *view* on their state. This requires the same level of dynamicity regarding field access which we assume for method dispatch.

Additionally, we need to maintain the *field invariant*, which can be achieved using at least the following two approaches:

1. *Strict:* An *Edit Transaction* can only add a field to a class if it neither exists in the inheritance path to the top nor in the subtree of subclasses of this class, except it also contains a field removal for the conflicting field (e.g. it represents a full pull-up field refactoring).

2. *Relaxed:* Field addition and removal can stack, i.e. existing fields may be re-added and shadow the field with the same name, while non-existing fields can be deleted without causing an effect. More elaborate edge-cases occur when multiple *Edit Transactions* are stacked, which we will deal with in section 2.4.

For this work, we opt for the second, *relaxed* model, as it allows a more permissive composition of *Edit Transactions* at run-time instead of forcing the developer to resolve all conflicts before a change is complete. For the change to be persisted in the base system, we need to make sure the field invariant holds eventually, i.e. by actually removing conflicting fields in subclasses if the environment requires this.

**Changing the class hierarchy**   An *Edit Transaction* should record the change to the superclass of a class. As with field changes, we need to re-establish the field invariant in either of the ways proposed for field changes. Concerning methods, we require a late-bound super call that enables us to request the method resolution order from the active *Edit Transactions*. Depending on how far-reaching the required change to the method resolution mechanism is, we need to take care of the hierarchy invariant too, e.g. by keeping deleted classes alive as long as classes in different *Edit Transactions* inherit from them. Often this is done by the runtime environment anyways.

**Altering the environment**   Adding, deleting, and renaming classes are environment-changing operations, i.e. concern a partial mapping $E(n) = c$ from a (global) class name $n$ to a (first-class) class $c$. Instances maintain their class, as they should refer to classes by identity and not by name. In case of deletion, instances continue to exist and behave like the last known version of their class (class preservation invariant), but no new instances can be created by referring to the class name since $E(n) = $ nil. An *Edit Transaction* contains a *differential* environment $E_\Delta$, such that the environment $\bar{E}$ presented to the control flow after activation satisfies $\bar{E}(n) = E_\Delta(n)$ if $n \in \text{domain}(E_\Delta)$ else $E(n)$.

## 2.4 Activation, Staging, and Dependencies

An *Edit Transaction* can be staged and activated independently. First, we describe a scenario where only a single *Edit Transaction* is maintained:

**Write access**   If there is a *staged Edit Transaction*, each change to any meta-object is being captured by this *Edit Transaction*. Underlying meta-objects of the base system remain untouched. If no *Edit Transaction* is staged, the base system is adapted as usual. Technically, Edit-Transaction-aware tools require read access to the source code and write access to source and executable representation of meta-objects, hence a staged *Edit Transaction* is being used to provide the work-in-progress snapshot on the source code without affecting execution until activated.

**Read access**   While an *Edit Transaction* is *active*, all meta-objects appear as changed in the respective *Edit Transaction*. E.g. method resolution will dispatch to the changed method rather than the version present in the base system, and newly added object state is accessible. Only unchanged meta-objects are re-used from the base system.

**Stacking *Edit Transactions***   We do not restrict our approach to having a single *Edit Transaction* overlaying the base system. We allow the introduction of a new *Edit Transaction* layered *on top* of a stack of existing ones for both *staging* and *activation*. Only the most recently staged *Edit Transaction* captures changes. At run-time, multiple stacked and active *Edit Transactions* appear as if the changes they contain are applied to the base system in their order of activation, which means that the meta-object versions in the more recently activated *Edit Transaction* have precedence over the next lower level.

It may be the case that different *Edit Transactions* modify the same meta-objects, e.g. a newer *Edit Transaction* adding class members deleted in a previous *Edit Transaction*, or removing previously added members, as Transaction 2 in figure 2(1 – 2) does. This creates *dependencies*.

As long as all dependencies are fulfilled by reproducing the staging order during activation, the composition is safe and indistinguishable from having the changes applied directly to the base system. However, we wish to retain full flexibility in activating, deactivating, re-ordering, staging, and un-staging *Edit Transactions* to develop, test, and combine them in isolation. This may lead to unsafe compositions, such as the one in figure 2 (3) where the deactivation of Transaction 1 broke dependencies Transaction 2 relied on.

Similar to the design choices we were faced with while maintaining the field invariant, we consider two models:

1. *Strict:* Dependencies are tracked and enforced at activation time, i.e. *Edit Transactions* with unsatisfied dependencies cannot be activated or force atomic co-activation of their dependencies.

2. *Relaxed: Edit Transactions* may activate as any subset and any order and conflicts emerge as run-time errors, are resolved using a specific protocol, or have no effect at all if the dependency is irrelevant.

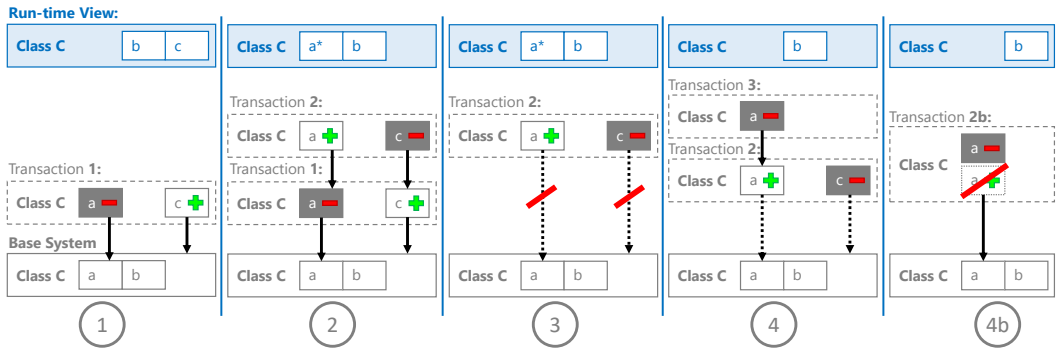

■ **Figure 2** Dependencies and conflicts on an example class. The top shows the run-time view on the class at different stages, the base system is at the bottom. (1) Transaction 1, deleting field *a* and adding *c*, is activated. (2) Transaction 2 re-adds *a* (*not the same a*, but a new field with the same name), deletes *c*, implicitly creating dependencies (solid arrows). (3) Transaction 1 is deactivated or discarded, breaking dependencies of Transaction 2. Our mitigation strategy shadows *a* at this point and the deletion has no effect. (4) The doubly-added *a* is removed by a third *Edit Transaction*, both original and shadowing *a* are invisible now since lookup stops at the deletion. (4b) Deletion takes place in the same *Edit Transaction* as addition, overwriting the add operation since both are mutually exclusive per *Edit Transaction*, and hiding the old *a*, since top-down lookup stops with deletion.

Since changes to methods, fields, and inheritance heavily rely on late-bound, dynamic dispatch, and dependency management adds a significant complexity, we opt for the relaxed model and resolve conflicts at run-time.

**Conflict Resolution**  We decide to use a conflict resolution strategy of least surprise. The most recently performed action in the most recently activated *Edit Transaction* which modifies a conflicting meta-object should "win" to save programmers from having to consider more changes down the stack.

Adding an already existing meta-object, as in figure 2 (3) will be resolved by *shadowing* the old meta-object. This way, Transaction 2 has the same behavior as when the field had been removed in between, and does not suddenly expose old data in the former field *a*. Methods are less of an issue, as they do not hold instance data.

Deletions will stop the top-down lookup for a named meta-object immediately. Hence, it does not matter if a meta-object did not exist before, like member *c* in figure 2 (3), or has been added multiple times, as *a* in figure 2 (4), where an already shadowed *a* is being deleted by Transaction 3.

Adding and deleting meta-objects are *mutually exclusive* operations per *Edit Transaction*. The last action performed in a staged *Edit Transaction* is the one effective during activation. For example, in figure 2 (4b), programmers might want to delete the field *a* again in the same *Edit Transaction* instead of a separate one, thereby discarding the add operation as if it never happened. The overall behavior stays the same: only member *b* is visible at run-time.

## 2.5  Scoping

An important feature of *Edit Transactions* is their capability to change their activation scope over time. We propose the following types of scope changes:

**Block-local activation**   A specific set of *Edit Transactions* is being activated for the duration of a code block. After the control flow exits the block, the previous activation state is restored. This code block may, for example, run tests or execute a user-provided code snippet under consideration of the changes accumulated in the active *Edit Transactions*.

**Thread-local activation**   When an instance of the currently edited program is running concurrently, changes can be pushed to the active control flow (thread). The other activity makes use of the changed meta-objects as soon as possible to maintain immediate feedback, but as late as necessary to keep the change *atomic*. Different considerations on a suitable point for *Edit Transaction* activation are discussed in section 2.6. Thread-local activation and deactivation permanently override block-local activation, e.g. exiting a block where certain *Edit Transaction* has been activated does not automatically deactivate this *Edit Transaction* when it has been thread-locally activated during the block.

## 2.6  Concurrency Control

Since we require the changes captured inside an *Edit Transaction* to emerge atomically, we need to take special care of pushing changes to other control flows, as the following examples demonstrate:

1. Given three methods $M_1, M_2$, and $M_3$, consider a thread having the call stack $[M_3, M_2, M_1]$.[1] We now have an *Edit Transaction* containing changes to $M_2$ and $M_3$. While control flow is in $M_3$, we can replace neither $M_3$ nor $M_2$. Even if control flow returned to $M_2$ we should not replace $M_3$ since there is the chance that $M_3$ is called again and would dispatch to the newer version while the old $M_2$ is still executing, thereby violating our requirement that changes come into effect atomically.

2. Consider $M_1$ performing an iteration that repeatedly calls $M_2$. If we update $M_2$, that iteration changes behavior. If $M_1$ is a stepwise update function in a grid-based simulation and iterates over grid cells, we would simulate a part of the grid with an old update and the rest with a newer one.

   These two observations give rise to consistency levels that decide at which points it is safe to activate an *Edit Transaction*:

**Method-level consistency**   Control flow can switch to the changed meta-objects as soon as the next method is being dispatched. This only guarantees that executing

---

[1] Top of stack is left to stay consistent with cons-style list notation, i.e. $[TOS|Tail]$

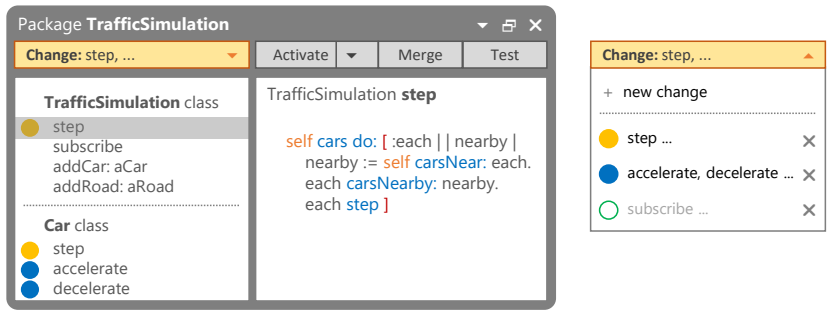

■ **Figure 3** Schematic program editor with program structure and code editor. *Edit Transactions* are indicated by color, the field in the top-left corner indicates which *Edit Transaction* is staged. *Edit Transaction* activation and tool support inside the *Edit Transaction* can be controlled from the editor. On the right, the drop-down menu hidden behind the staged *Edit Transaction* indicator allows to activate, deactivate, add, and delete *Edit Transactions*.

methods on the call stack continue to execute their old code (which is the default behavior in many live programming environments), but recursive methods can already dispatch to newer versions of themselves.

**Reentrant consistency**   Changes get applied from the moment where the call stack does not intersect with the set of changes, i.e. the moment when the last method modified by the *Edit Transaction* returns, all changes are installed and become effective. This would address the first example above. Reentrant consistency is analogous to the *passiveness* criteria that are part of the notion of *quiescence*[14] and *tranquility*[29], except nodes are methods here.

**Manual consistency**   The second example above depends on domain-specific semantics. It is up to the programmer to place explicit *atomic regions* in the code using a language feature or library call. If consistency level is set to *manual*, activating *Edit Transactions* are delayed until the atomic region is left.

**Parallel activation**   Since a program can run multiple threads and we want to give the programmer the chance to atomically activate an *Edit Transaction* across a subset of (or all) threads, we need to install a barrier once a thread reaches updateable state (depending on the consistency level). The last thread entering this barrier causes the *Edit Transaction* to activate on all threads and release the barrier.

## 3    Tool Support

### 3.1  Program Editor

We suggest a slight extension of existing editing tools of the environment, like illustrated in figure 3. The following actions should be possible within a program editor:

- Create, stage, and un-stage *Edit Transactions* to control which in-progress view on the program is seen and modified by the editor.
- Activate and deactivate a selection of (or all) staged *Edit Transactions* and extend the activation scope to a running program or the complete environment. Meaningful scopes can be proposed by the tool or added by the programmer.
- Discard an *Edit Transaction*. It will deactivate globally and changes are lost.
- Merge the selected *Edit Transaction*, either with the *Edit Transaction* one below or with the base system.

### 3.2  Testing

One of the most important mechanisms to obtain feedback on program changes is unit testing, either initiated manually or automatically triggered by changes to the program. Therefore, we propose a tight coupling with a test runner.

**Synchronized test runner**    Opening a test runner from the current editor links the *Edit Transaction* activations of both tools, i.e. the test runner tests exactly the perspective taken by the editor. If the test runner is set to *auto-test*, each change in the corresponding program editor will cause tests to run. Analogously, debugging a failing test will happen in the scope of the currently selected *Edit Transactions*, i.e. code changes done in the debugger or test runner are collected by the same *Edit Transaction* as in the program editor.

**Automating *Edit Transaction* actions on passing or failing tests**    When all tests pass, we suggest to let the programmer select if the successfully tested *Edit Transaction* should activate automatically or at least activate a shortcut or button to activate or merge. This would require coverage of the changes captured by this *Edit Transaction*, otherwise the test results give no feedback on these changes.

### 3.3  External Changes

Tools capable of loading external changes, such as version control software integrated in the live programming environment, might be set up to load their changes into a separate *Edit Transaction*. This way, changes can be reviewed and tested within a program editor or test runner. The new behavior can subsequently be activated globally with the option to quickly undo the changes in case of problems.

## 4  Implementation

We describe an implementation for *Edit Transactions* in the class-based object-oriented live programming environment Squeak/Smalltalk. We make use of the existing meta-object protocol to implement *Edit Transaction* behavior without changes to the virtual machine.

### 4.1  Capturing Method Changes

**Augmented method dispatch**   In Smalltalk, a class maintains a *method dictionary* mapping message selectors to `CompiledMethod` objects. A `CompiledMethod` contains the executable form of a method, i.e. its bytecode and layout of the stack frame, and is interpreted by the Smalltalk virtual machine.

If the object encountered after method lookup is not a `CompiledMethod`, the message `run:` selector `with:` arguments `in:` receiver is sent to this supposedly method-like object with the original message selector, an array containing the original arguments, and the receiver passed as arguments.[2]

This *object as method* strategy allows us to add a new dimension of dispatch: Instead of a `CompiledMethod` object, our implementation uses `MultiVersionMethod` objects, which internally store a mapping from *tags* to `CompiledMethod` objects (see figure 4). The tags identify the *Edit Transactions* that have changed this particular method. The `run:with:in:` method asks the current thread (`activeProcess` in Smalltalk) which *Edit Transactions* are active, selects the tag attributed to the top-most *Edit Transaction* that has changes in this method, and forwards the call to the associated `CompiledMethod` version.

**Trade-offs**   When it comes to live programming environments, we always prefer immediacy of feedback over long-term computational efficiency during development. This trade-off presents itself in our method dispatch implementation, as *Edit Transaction* activation and deactivation is as instant as adding and removing an element from a list, while we accept a per-call overhead. With modern virtual machines that support just-in-time compilation, this overhead would be eliminated if the activation stack stays constant for several thousand calls [18].

**Changing methods**   In order to capture method changes, we adapt the `compile` methods of the *metaclass* `ClassDescription` to handle `MultiVersionMethod` objects in the method dictionary. If a `CompiledMethod` in the method dictionary is being replaced, the metaclass moves it to the `MultiVersionMethod` under the `#base` tag identifying the case when no *Edit Transaction* is active and adds a new `CompiledMethod` indexed by the top-most active *Edit Transaction*. Every subsequent change to the same method within the same *Edit Transaction* overwrites this `CompiledMethod`.

---

[2] Note that this is similar to Python's \_\_call\_\_, except that original receiver and selector are additionally passed through by the runtime.

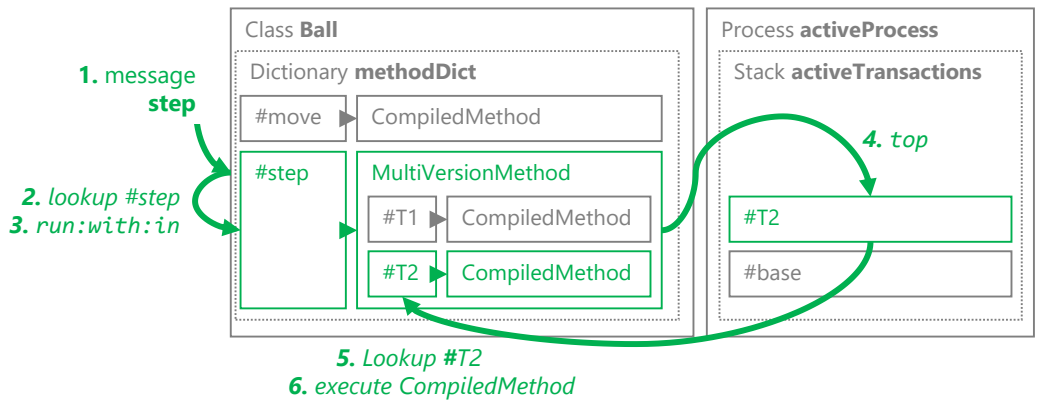

**■ Figure 4** Method dispatch with *Edit Transactions* as additional dispatch dimension and the data structures used in Squeak/Smalltalk. In this example, one of two *Edit Transactions* is active and the step method has been modified in both. Arrows indicate the order in which data is being consulted to retrieve the actual `CompiledMethod` containing the executable bytecode in reaction to receiving the step message.

Staging is managed by program editing tools. They are expected to pass the identifier of the *Edit Transaction* that should capture the recompiled method as additional argument to the `compile` methods.

### 4.2 Capturing Field Changes

**Private fields and single storage**  In Smalltalk, field access is restricted to methods of the class and statically bound to a specific memory offset in the instance at compile time. Adding, removing, renaming, or reordering fields would require instances to migrate to a new memory layout, possibly losing information, which would be a side-effect to avoid. Another Smalltalk-specific peculiarity occurs if a field is being removed but still referenced by a method. In this case, the field will be removed from all instances and redirected to a *single storage*, i.e. all instances read and write the same value. There is one single storage per class and name combination.

**Adding and aliasing fields**  Whenever a field is added, the currently staged *Edit Transaction* records the respective field as modified and thereby indicates to the compiler to use dynamic field lookup on this particular field. For each new field, a unique *alias* is assigned. Aliasing is used to distinguish between fields with the same name added in different *Edit Transactions* and realizes shadowing in case of conflicting activation.

This allows us to emulate the Smalltalk behavior of losing the value attached to a field if it is being removed and re-added, but at the same time maintains our *information preservation invariant*, i.e. a control flow where the *Edit Transaction* removing the field is not active does still have access to the old field values.

Old methods not modified cannot know about the newly added field and therefore remain untouched, while new methods using this field get compiled with dynamic field lookup.

■ **Listing 1**  Edited and generated code after introducing dynamic field lookup in an Edit Transaction. Instance variable *a* has been added inside a staged Edit Transaction before.

```
1  MyClass >> inc   "Code edited by and displayed to programmers"
2      a := a + 1
3
4  MyClass >> inc   "Generated code in a staged Edit Transaction"
5      Storage at: self set: #a to: ((Storage at: self get: #a) + 1)
```

**Dynamic field lookup**  In our implementation, the `compile` methods replace static binding by dynamic lookup when an instance variable occurs in an *Edit Transaction* using an upfront transformation of the parse tree. The dynamic lookup is a call equivalent to `Storage at: self get: #name`, with `#name` being the symbol representing the field name. Dynamic writing is `Storage at: self set: #name to: value`. An example is shown in listing 1. The content of these dynamic fields is managed by the `Storage` and not necessarily co-located with the instance.

During lookup, the topmost *Edit Transaction* is asked for the modification state of the field in the respective class. In case of a removed field, lookup proceeds with the superclasses inside the same *Edit Transaction*. If any field definition is encountered, its alias will be retrieved. If no addition or removal has been identified, lookup repeats with the underlying *Edit Transactions* and eventually the base system. If an alias was successfully retrieved, it is combined with the instance's identity and used as key in the `Storage` object to read or write the value, otherwise the corresponding single storage location will be used, which is identified by class name and field name.

**Deleting fields**  For deletion, we consider two cases: Deletion of a field which has been added within the same *Edit Transaction* will just remove the field from the *Edit Transaction* again, methods already compiled with dynamic lookup for this field will be recompiled and the `self` replaced by the reserved identity for single storage. Any new method referring to this field results in a compile-time error. In the second case, the removed field was in the underlying base system, which means, we need to recompile even methods originally not in the *Edit Transaction* to refer to single storage when accessing this variable and put their recompiled version into the *Edit Transaction*.

All in all, instance state changes referring to not-yet committed field changes are kept separate from the instance seen by the base system, yet we can emulate a different instance layout dynamically at control-flows where an *Edit Transaction* is active (see figure 5).

### 4.3  Scoping *Edit Transactions*

Methods and environments dispatch according to a stack of active *Edit Transactions*. This stack is part of the `Process` class, which represents the concept of threads. This association of transaction stacks with Processes allows programmers to debug code of transactions in separate processes. This is of particular interest in combination with the Morphic framework of Squeak/Smalltalk [16]. In case of an exception in the

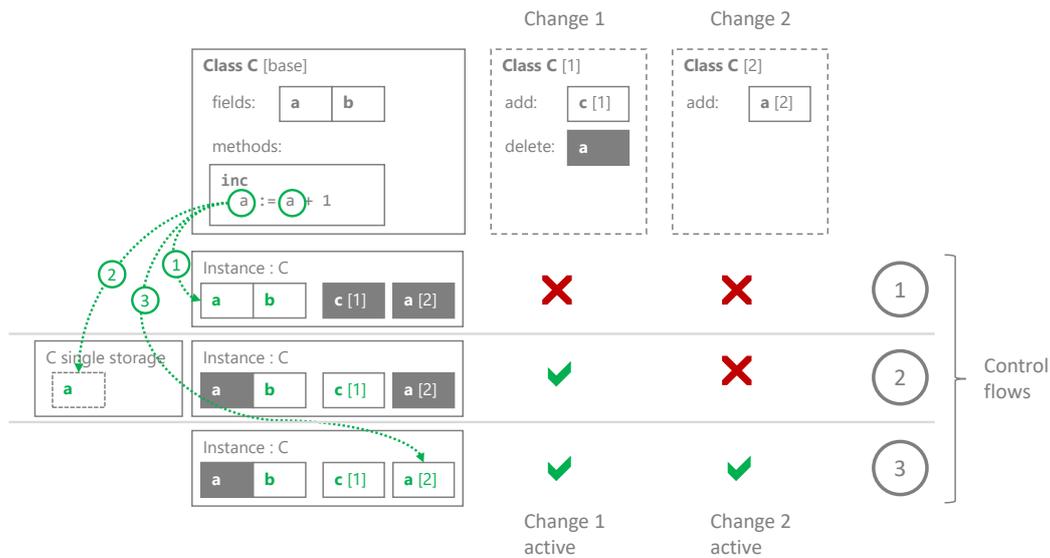

■ **Figure 5** Field dispatch with *Edit Transactions*. The same instance of class *C* is shown in three different control flows having none, one or two changes active. The variable *a* referred to by the inc method is being dispatched as shown by the dotted green arrows in the respective control flows. Numbers in square brackets indicate aliasing and are not part of the identifier name. Shaded fields are not visible.

UI process, the process is suspended and the environment opens a debugger for it. Additionally, a new UI process is spawned which has an empty transactions stack. If the exception was caused by code in one of the active transactions, the new process will thus not throw the exception again.

Block-local activation is implemented by the `BlockClosure` class, so that given code like [self step] inTransactions: #(tag1 tag2) pushes the *Edit Transactions* corresponding to the given tags on the stack of `Process activeProcess`.

### 4.4 Implementing Consistency

In our implementation, method-level consistency is trivial: Just adding and removing *Edit Transactions* from the activation stack at the target `Process` instance takes effect at the next method dispatch.

Implementing reentrant consistency can be achieved by creating a `BlockClosure`, which would set the activation stack if evaluated. The target activation state is captured within this block closure. If the target `Process` instance is the active process, we can retrieve the current call stack context using `thisContext`, otherwise, we request the `suspendedContext` of the process. These context objects represent our stack frames that need to be scanned for any method involved with the *Edit Transactions* to be activated. Traversal is done by recursively following the `sender` variable. The sender of the topmost context from which on no conflicting method has been found is stored and the block closure converted to a context using the `asContext` method. The stored

■ **Listing 2**   Example code modified during execution

```
1   Simulation >> mainloop
2     [self running] whileTrue: [self step]
3
4   Simulation >> step
5     self balls do: [ :each |
6       each move]
7
8   Ball >> move
9     self pos: (self pos + self speed)  ⇐
```

■ **Listing 3**   Modified example code in an *Edit Transaction*

```
1    Simulation >> mainloop
2      [self running] whileTrue: [self step]
3
4    Simulation >> step
5      self balls do: [ :each |
6        each bounce; move]
7
8    Ball >> move
9      self pos: (self pos + self speed);
10       speed: (self speed * self drag)
```

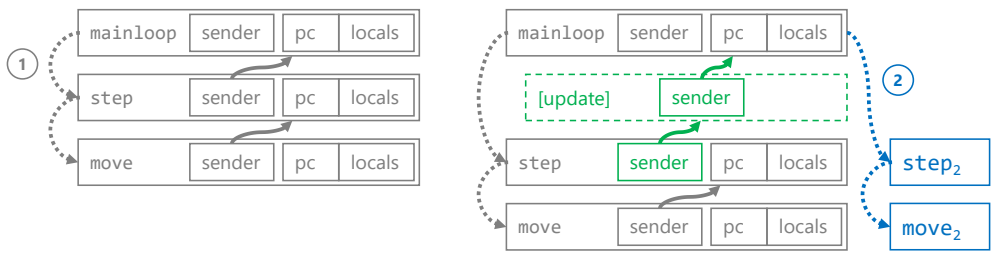

■ **Figure 6**   Asynchronous *Edit Transaction* activation via stack-frame insertion. mainloop repeatedly calls step, which in turn calls move (1); the thread's call stack halted at listing 2, line 9 (⇐) is shown (the sender field refers to the context to be returned to). An *Edit Transaction* updating both step and move (listing 3) should be activated, which causes the temporary context [update] to be inserted on top of step. If step returns, [update] activates the change and returns to mainloop, such that the next call (2) executes the updated methods from listing 3.

context's sender is set to the result of this call, whose sender is in turn set to the stored context's original sender, thereby intercepting the return of this context. Since the intercepted return may also return a value, we need to instruct the closure to pass through this value by returning thisContext at: 2 (the stack frame location where the last return has been stored). The result is illustrated in figure 6.

Manual consistency is provided by the method Process activeProcess **update**, which simply overwrites the current stack of *Edit Transactions* with a pending stack of *Edit Transactions*. Instead of the active process, this method can be invoked externally on any process, bypassing consistency guarantees.

### 4.5   Tool Support

We use the data-flow based tool-building framework Vivide [25] to construct a browser similar to those known from Smalltalk environments.

The program editor as illustrated in figure 7 is constructed in a way that it might edit any *Edit Transaction* and view the source code through an arbitrary subset of *Edit Transactions* without actually running inside any active *Edit Transactions* itself. The outer left user interface elements give the programmer control over the existing

**Edit Transactions**

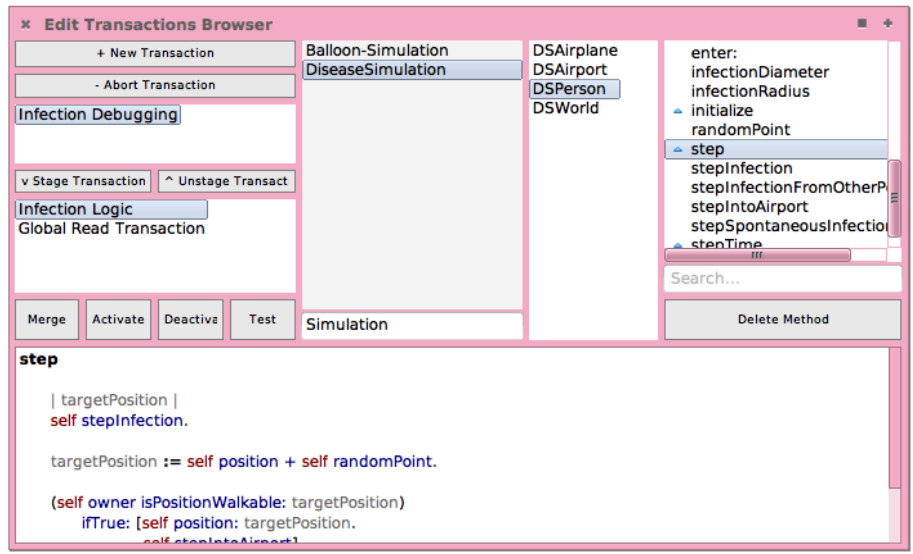

■ **Figure 7**   A screen shot of the *Edit Transactions* browser showing the view on the system as described by the "Infection Logic" transaction. The outer left user interface elements allow programmers to control the state of transactions. All changes applied in the code pane at the bottom are captured in the topmost transaction in the lower list on the left (in this case "Infection Logic").

transactions in the system (upper list) and the staged ones (the lower list). The buttons in between the lists state or un-stage transactions. These transactions are selected for all transaction browsers in the current user interface process. To globally activate transactions, developers can use the "Activate" and "Deactivate" buttons. The "Test" button executes any tests in the package selected in the second column in the scope of the staged transactions. If a change to a method or a class definition is saved, it is applied within the topmost *Edit Transaction* in the lower transactions list.

## 5  Case Study

To investigate the potential impact of Edit Transactions, we conducted a preliminary exploratory case study [21, 22]. The *objective* was to determine pointers for improvement of the design and relevant aspects for a thorough evaluation. The *case* we studied is that of adding a new feature to a simulation. The same feature was once implemented using standard tools and once using Edit Transactions. The concrete *research question* we wanted to answer was how our approach changes the programming practices and experience. To *collect data*, we took video recordings and manually logged when debuggers were displayed, the type of exception which caused the debugger to open up, and how the underlying issue was resolved. We also noted any notable programming practices based on subjective judgment (e.g. which tools were used together). The case was executed by one of the authors who has more than five years of experience using Squeak/Smalltalk and is familiar with the pitfalls of a live programming environment. Further, the programmer also took part in the development of the tool and thus there was no additional training.

In the following sections, we describe the application and task used in the case study, the observations for both variations, and the impact of *Edit Transactions* on the responsiveness of the environment.

## 5.1 Case: Disease Spreading Simulation

The case was conducted with a simulation of the spreading of diseases between people. It is based on Person objects doing a random walk on a world map. Their infection status is modeled as a Boolean instance variable called infected. Persons can either get a spontaneous infection or they can get infected by an infected person nearby. The simulation is implemented in the *Morphic* framework [16] for developing graphical applications. The framework provides the *stepping* mechanism which periodically calls the step method on all Morph objects implementing it. The main part of the simulation logic is implemented in the step method of the Person class. The simulation was initialized with 750 Person objects. In both variations, the task was to modify the source code while the simulation was running next to the programming tools.

**New Feature: Recovering from Various Infections**   The task for the case was to add recovering to the simulation. The recovery should take different amounts of time depending on the type of infection. This implied a number of technical changes which needed to be implemented. These technical adaptations are the assignment to be fulfilled. First, the infection status has to change from a Boolean variable to a reference to Infection objects. Accordingly, all methods assuming the infection status was a Boolean value have to be adapted. Also, basic infection classes have to be added. To bring the logic together, we must add a call from the step method of a Person to its infection and we have to add methods to handle a full recovery.

## 5.2 Workflow without *Edit Transactions*

The first type of error is access to uninitialized or lost state. This first occurred, when we replaced the infected instance variable with the infection variable. By doing so, we accidentally destroyed the information on whether a person is infected. Thus, immediately after saving the changed class definition, the environment opened several debuggers, one for each Person object trying to access its obsolete infected variable. To control the flood of incoming debuggers, we closed the simulation. Re-adding the instance variable would not have helped at this point as the state was already lost. This kind of error occurred again, when we added the recoveryTimer instance variable to the infection class and used it in a method. To not loose the simulation state, we tried to initialize the old objects by executing a code snippet. However, as debuggers capture the keyboard focus when they become visible and debuggers were appearing with a high frequency we could not edit any text. To resolve the issue, we had to restart the simulation. The second class of errors occurred when we used messages in a method which were not implemented yet. When this type of error occurred while the simulation was running, again a lot of debuggers opened. Often, we could resolve the issue without closing the simulation by quickly removing the call to the unimplemented method.

We made the following observations on the programming practices. First, most of the time two class browsers were open. The second browser was used whenever there was a change in the first browser which would cause the simulation to crash because of unimplemented messages. In such cases, the programmer switched to the second browser to implement the required messages and only then saved the method in the first browser. Regarding the implementation strategy of new behavior, we observed that the programmer first implemented a set of interacting methods and only in the end added the calls triggering the set of methods to the running methods. Also, we observed that the programmer pro-actively restarted the simulation to create a consistent application state through forcing the execution of initialization methods.

### 5.3 Workflow with *Edit Transactions*

When using *Edit Transactions*, the first type of error we encountered was that of uninitialized instance variables again. In one instance, we activated a *Edit Transaction* which contained the replacement of the infected variable. This time only one debugger opened up and the simulation kept on running. As the debugger does currently not support evaluating code in the context of the process being debugged, we could not initialize the uninitialized objects directly from within the debugger. We resolved this by adding a method to Person which initializes all Person objects and then added this method to the stepping callback. We activated the *Edit Transaction* for a short period of time and afterwards removed this method again. Beside these state-related exceptions, several debuggers opened up due to messages which were used but not yet implemented. All of these cases only opened one debugger window and kept the simulation running.

We observed the following practices in combination with *Edit Transactions*. In general, before activating a transaction, the programmer edited several methods and reviewed the change. The activation of transactions was often only for a short period of time to see whether the change was working at all. Further, during the whole session only one class browser was used, although there was the option to open more browsers. Most of the time only one transaction was used to implement the current task. The programmer used one transaction to implement the switch from representing an infection as a Boolean value to an object. After finishing and merging this change, the programmer used another transaction to implement the recovery logic. During the recovery implementation, the programmer temporarily added a transaction which added logging statements displaying the distribution of infections in the population to validate the recovery logic. On several occasions, the programmer applied the method to initialize added instance variables as described for the first error type. This was possible because of the atomic activation of a transaction which includes added state as well as the added call to a temporary initialization method.

**Responsiveness of the environment**   In a self-sustaining live programming environment, any penalty on the execution speed of the application can also affect the development tools. We noticed that the development tools became less responsive when using Edit Transactions.

While a thorough quantitative investigation of the execution times is beyond the scope of this case study, we briefly investigated the perceived degradation of responsiveness. Running the simulation with an open class browser resulted in a rendering frequency between 18 and 26 frames per second (fps). Adding a field to the class `Person` added no penalty except for a short drop in fps for several seconds. When running the simulation with edit transactions the frame rate dropped. Creating a transaction which adds one field caused the frame rate to drop down to 8 to 14 fps. This happens as by adding a field, the transaction adds dispatch logic to almost all methods of the `Person` class. Activating the transaction did not affect the frame rate any further. Also adding a second transaction had no further effect.

These observations are supported by preliminary benchmarks showing the potential overhead for method calls and state access (for details see appendix A). The current measurements suggest a constant slow-down of a factor of 40 for calls to methods which are modified in a transaction in the system independent of the number of active transactions. Further, there is a slow-down of factor 36 for a method accessing one instance variable added in a transaction up to a factor of 2288 for a method accessing ten instance variables.

*Edit Transactions* are a concept designed for live programming environments. So, while the observed degradation of responsiveness did not massively disturb the live programming experience, future implementations of *Edit Transactions* have to take the impact on the performance of the system into consideration.

## 6   Discussion and Future Work

### 6.1  Limitations of the Implementation

The concept of *Edit Transactions* has not been fully implemented yet in our prototype. To conduct the case study, we have implemented support for all meta-object changes except changes to the class hierarchy. Super calls are statically bound at compile time in Squeak/Smalltalk, leaving almost no chance to dispatch the super call dynamically with regard to the activation without changing the virtual machine itself. Additionally, there is currently no graphical way of extending the scope to specific threads. This needs to be done programmatically.

The workflow itself could be streamlined by introducing keyboard shortcuts, e.g. CTRL + SHIFT + S to save to a new *Edit Transaction* instead of the normal CTRL + S to save the method to the base system or currently staged *Edit Transaction*.

## 6.2 Research Questions for Quantitative Evaluation

We presented a case study to gain initial insight into a live programming workflow with customizable change granularity and scope, but no hypotheses have been tested in this design and engineering phase of our research yet.

Of particular interest for us is the question, whether test-driven methods benefit from *Edit Transactions*, since we suspect a mixture of immediate feedback on test runs and a safety-net given by *Edit Transactions* to reduce the number of errors and setbacks. Testable null hypotheses would include the assumption of observing an equal number of errors caused by an implementation task, equal degrees of task completion, and equal time to completion compared to a fully immediate workflow. Additionally, we are interested whether programmers would spend a higher proportion of their time coding alongside a running program, which would indicate a higher continuity of feedback and thereby increase *liveness* according to multiple perspectives on what liveness means [4, 26, 27]. Negative results on these hypotheses might suggest that unmediated emergence of changes, despite being at continuous risk of breaking the program, is still the best feedback so far.

## 6.3 Related and Future Language Concepts

Some solutions we implemented can be viewed as isolated language concepts which we prototypically introduced. Some of them resemble ideas from literature, while others could be proposed as stand-alone concepts in future work with *Edit Transactions* being an example use case for them. The following subsection summarizes these insights.

**Shape-shifting instances**   By making field access dynamic and dispatching the lookup through a stack of *views*, we can present the same object differently to different control flows. Fields can be hidden, but information is being preserved. We disambiguate between fields with the same name but different meaning by transparently aliasing them depending on the active view. Views on instances are composable. While the concept is quite specific, it could extend context-oriented programming [10] in a way that is similar to the partial objects proposed in $L$ [11], except we do not deal with explicit visibility but resolve conflicts as they emerge during composition (see section 2.4).

**Dynamic re-classification**   Dynamically changing the inheritance hierarchy depending on active *Edit Transactions* has been addressed recently using just-in-time inheritance [5]. This particular approach persistently changes the representation of objects to adhere to a different superclass precedence in multiple inheritance scenarios. With *Edit Transactions*, in comparison, a class can be switching to any new superclass independently of a multiple inheritance scheme, and the new perspective should only be transiently emulated within certain control flows. We see some future work in exploring dynamic and scoped reclassification of objects.

**Polymorphic identifiers**  The concept of *polymorphic identifiers* [30] generalizes left-hand values and associates them with schema handlers that deal with how a value is written to read from the storage location pointed to by the identifier. If identifiers for methods, instance variables, and global variables (e.g. classes) were polymorphic, we could easily install schema handlers capable of dispatching through a stack of active *Edit Transactions*. Also, write access to meta-objects could be captured in a modular fashion by replacing or extending schema handlers.

**Mirrors for *Edit Transaction*-aware reflection**  At the moment, using reflection capabilities will bypass any abstraction we created with our *Edit Transaction* implementation; there is no reflective API providing an "inside view" given a stack of active *Edit Transactions*. As an example, iterating over instance variables does not reflect changes by active *Edit Transactions*, since they are not co-located with the object. We could adapt each reflective method, which would be cumbersome and not modular.

A more principled reflection facility are *mirrors* [3], where reflection capabilities are provided by a separate meta-object model. Since we might want to present a program using reflection an "inside view" of the system, we could implement specialized *Edit Transaction*-aware mirrors for normal reflection. The view dynamically generated by activating *Edit Transactions* would affect corresponding inside mirrors as well. Interesting future work might emerge from the idea of switching between inside and outside view of an abstraction like *Edit Transactions*, e.g. for debugging purposes.

## 7  Related Work

**Version Control Systems**  Version Control Systems (VCS) like Git,[3] Apache Subversion,[4] or Mercurial [5] are based on similar ideas, but their usage scenarios are different. One analogy that can be drawn is the correspondence of active *Edit Transactions* to revisions or commits in a VCS, deactivation to reverting a commit, and a staged but not yet active *Edit Transaction* to the working copy. The stash operation known from Git is similar to un-staging an *Edit Transaction*. Extending the scope of an *Edit Transaction* activation can be regarded as pushing (in the Git/Mercurial sense) that commit to a remote copy, such as a continuous integration server, or a deployment server where the jointly committed and tested behavior emerges in the running application. Activating *Edit Transactions* out of order resembles doing a *rebase* operation on the respective commits. A major difference is the fact that *Edit Transactions* are explicitly designed to forget history once they are merged. Future work could include synchronizing an *Edit Transaction* merge with a VCS commit, which might yield "natural" commit sizes. The other way around is possible in our implementation, since pulling a commit can do all necessary recompilation inside a staged *Edit Transaction*, thereby allowing to explore this commit in the running program.

---

**Change-oriented Programming**    Change-oriented programming promotes the idea of making changes to the system description a first-class citizen accessible to the programming system and programmers alike [7, 23]. Thus, modifying a system becomes a matter of applying changes to the system description. As a result, the programming system or the programmer can reflect and work with changes, such as grouping them or replaying them in other settings [7, 17]. *Edit Transactions* combine this idea of reified changes with dynamic scoping. However, while change-oriented environments promote interactions with changes, *Edit Transactions* are a necessary mean to allow for an improve live programming workflow.

**Changeboxes**    As an instance of change-oriented programming, Changeboxes provide an object model representing changes as first-class instances in the environment [6, 32]. Part of this model is an extensive change specification, which encodes primitive actions (e.g. define, rename, and remove), composite actions (e.g. pull-up method and other refactorings) and their corresponding targets in the program (e.g. a class, method, or field). In order to scope a Changebox to a particular programming session and develop and run code inside a specific version, a thread-local context with the current Changebox is involved in method dispatch and environment lookups. This implementation strategy served as basis for our *Edit Transaction* implementation. Changeboxes have similar tool support to *Edit Transactions*, such as a Changebox-aware editor, debugger, and test runner.

A major distinction between both approaches is the goal of making program changes and intent retroactively accessible with Changeboxes, while *Edit Transactions* are designed for temporary containment of future behavior. An *Edit Transaction* flattens the result of a composite change, and, in contrast to Changeboxes, existing objects are not bound to a particular version and can be dynamically accessed from the perspective of the currently active *Edit Transactions*.

**CoExist**    The CoExist project proposed an environment which creates a new version for every individual change [24] . The Co-Exist environment allows programmers to go back in time, run multiple versions simultaneously, and compare them not only as code, but also as running instances. Granularity can be modified retroactively by grouping multiple correlated changes, which can even be used as change sets in a version control system. On the one hand, programmers are freed from remembering to explicitly manage a *Edit Transaction*, on the other hand they cannot choose to defer changes from being applied instantly, only revert in case of failure. A combination of both scenarios gives rise to promising future work.

**First-class Contexts**    Wernli et al.[31] propose a more explicit way of updating a system at run-time. Their *contexts* do not only capture change to meta-objects but also how to migrate application objects between versions. Once an update is deployed, new threads start within the updated context, while old threads still operate based on the old meta-objects. Objects can exist simultaneously in multiple representations, the last updated representation invalidates representations from other versions, reading an invalid representation lazily synchronizes it from a valid one.

The concept allows a similar workflow to *Edit Transactions*, such as extending the scope of an update, and allowing multiple control flows to coexist in different versions while sharing instances. The synchronization protocol required for an update makes the concept significantly more expressive at the cost of having to implement synchronization and migration. For example, the semantics of a field in a class may change and a migration can update that field for all instances.

Another trade-off is migration of instances versus late binding of state. The former creates update overhead but minimizes run-time overhead, which is desirable for deploying an update to a production system. The latter, which we preferred in our case, trades per-lookup overhead for immediate feedback as discussed in section 4.1.

**PIE**   The personal information system (PIE) included mechanisms through which "descriptions of alternative software designs can be readily created and manipulated" [2]. The stated intent was to enable developers to compare design alternatives more easily. One of the central design principles also applies to *Edit Transactions*: "... there should exist a descriptive level at which objects can be described without actually affecting the objects themselves." Our staged *Edit Transactions* in the browser enable such a workflow of viewing various alternatives without affecting the behavior of the objects. At the same time, our intent is different as we focus on improving the live programming experience of editing a system while it is running. This leads to our requirements of maintaining consistency when activating a *Edit Transaction*.

**Development Layers**   A particular problem in collaborative, self-sustaining programming environments, such as Lively [13], is, that a breaking change to a development tool effectively breaks the tool for everyone, including the developer itself. The concept of *development layers* [15] allows to scope changes to tools by using *context-oriented programming* [10] and putting them inside a *layer*. A separate tool can now control when such layer is active. Development layers do not support back-merging into the base system and only handle behavior changes in a prototype-based environment (JavaScript), however, they served as inspiration for the development of *Edit Transactions* as dynamically activated changes in a live programming system.

## 8   Conclusion

In this work, we propose *Edit Transactions* to address the fragility that comes with immediate feedback when making changes to running programs.

Conceptually, *Edit Transactions* decouple the *adaptation* stage in live programming environments from *emergence*, allowing time and scope of emergence to be precisely controlled by programmers. We described the implementation of *Edit Transactions* and associated tool support in Squeak/Smalltalk, and discussed a number of insights regarding which parts of our implementation constitute programming concepts on their own or can be considered as applications of existing concepts.

From a programmers' perspective, *Edit Transactions* introduce a two-stage workflow into live programing in which programmers collect a number of changes to the *Edit*

*Transaction* before deliberately activating the new behavior in running programs. Our case study gives insights into a programming experience that constantly balances immediacy against the need for grouped changes and the associated safety net. Additionally, we retain some types of immediate feedback that work well in isolation, such as automated unit testing, and allow running programs to instantly fall back to a previous version whenever the newly activated change causes an error.

So far, *Edit Transactions* show potential to increase confidence in a change, prevent some classes of run-time errors of live systems, and eventually make live programming a more predictable and engaging activity.

**Acknowledgements**   We gratefully acknowledge the financial support of the Research School for Service-oriented Systems Engineering of the Hasso Plattner Institute and the Hasso Plattner Design Thinking Research Program. We thank Stefan Ramson for discussions of earlier versions of this submission.

## A  Micro Benchmarks for Method Call and State Access Overhead

In order to determine the overhead imposed by the new dispatch dimension we conducted micro-benchmarks. We conducted benchmarks measuring the overhead for method calls as well as for accessing instance variables. As transactions can be seen as COP layers, we used an existing set of COP dispatch benchmarks [1]. It consists of a benchmark class with 10 fields named field1 to field10. The class implements ten methods, which are named method1 to method10. Each method increments the value of the first n fields, so method3 accesses field1, field2, and field3. For measuring the impact of transactions, we add 9 transactions which each contain an implementation of method1 which behaves as one of method2 to method10. So, method1 with the active transaction2 also accesses field1, field2, and field3. Subsequently, these transactions are activated.

**Edit Transactions**

The state benchmarks are based on the same idea. There is another benchmark class which has one instance variable (called field1) and one method (called method1) which increments this field. Further, there are nine transactions which add several instance variables and a new implementation of method1 which increments all of these fields. The transactions are subsequently activated.

### A.1 Setup

We took then measurements of the duration of 1,000,000 method calls and took the median of the results. For the basic method call case we measured the duration of calls to method1 to method10. For the method call with transactions case, we measured the duration of calls to method1 with subsequently more active transactions with the top-most transaction providing the intended behavior. Similarly, for the state access with transactions benchmark, we measured the duration of calls to the method1 of the state benchmark class with subsequently more active transactions.

During the benchmark run, the garbage collector was disabled. All benchmarks were executed on the following system:

- Intel CPU i5-4690 @ 3.5 GHz, 4 Logical cores
- 7926 MB Main Memory
- Ubuntu 15.10
- Squeak 5.1 Update #16548
- Croquet Closure Cog[Spur] VM revision 201608171728

### A.2 Discussion of Results

Table 1 shows the benchmark results. The slow-down for the method call benchmark remains around a factor of 40 and seems independent of the number of active transactions. Notably, even without any active transaction, the mere presence of transactions containing changes for the benchmarked method already causes the slow-down. This is expected, as modifications to a method implies a MultiVersionMethod which adds extra overhead to the dispatch process. At the same time, it can also be explained by the dispatch algorithm implemented in MultiVersionMethods which checks the base system last.

The slow-down factor of the state access benchmark increases with the number of active transactions. As the benchmarks were executed on a virtual machine applying optimizations through just-in-time compilation, the observed increase might be attributed to these optimizations. Further benchmarks are required to clarify these aspects as well as any variations in the method call benchmark.

■ **Table 1** Benchmark results for method call overhead or state access overhead. The Without column lists the median durations for method calls without any active transactions, the Call column the median durations for method calls with active transactions, and the State column lists the durations for method calls with transactions adding state. The slow-down columns are relative to the basic durations.

| | Results in ms (Standard Deviation) | | | Slow-down | |
|---|---|---|---|---|---|
| No of Transactions | Without | Call | State | Call | State |
| 0 | 14 (6.26) | 579 (5.21) | 515 (9.85) | 41.36 | 36.79 |
| 1 | 15 (0.00) | 610 (2.74) | 5,854 (11.75) | 40.67 | 390.27 |
| 2 | 16 (0.42) | 654 (1.49) | 11,789 (41.95) | 40.88 | 736.81 |
| 3 | 17 (0.00) | 683 (1.23) | 18,328 (21.32) | 40.18 | 1078.12 |
| 4 | 17 (0.53) | 696 (1.52) | 21,989 (42.98) | 40.94 | 1293.47 |
| 5 | 18 (0.42) | 717 (0.85) | 27,653 (51.31) | 39.83 | 1536.28 |
| 6 | 20 (0.00) | 737 (1.51) | 33,179 (19.39) | 36.85 | 1658.95 |
| 7 | 21 (0.00) | 839 (0.82) | 38,830 (36.77) | 39.95 | 1849.05 |
| 8 | 21 (0.00) | 795 (1.08) | 44,368 (64.10) | 37.86 | 2112.76 |
| 9 | 22 (0.00) | 822 (3.20) | 50,339 (69.76) | 37.36 | 2288.17 |

## About the authors

**Toni Mattis** Is a doctoral researcher at the Software Architecture Group. His research interests include live programming and repository mining.

**Patrick Rein** Is a doctoral researcher at the Software Architecture Group. His research interests include live programming and programming systems.

**Robert Hirschfeld** is a Professor of Computer Science at the Hasso Plattner Institute at the University of Potsdam, Germany. His Software Architecture Group is concerned with fundamental elements and structures of software.